\documentclass[prd,twocolumn,notitlepage,nofootinbib,unsortedaddress]{revtex4-1}
\usepackage{amssymb}
\usepackage{amsmath}
\usepackage{graphicx}
\begin{document}
\title{Self-force as a cosmic censor}   
\author{Peter Zimmerman} 
\author{Ian Vega} 
\author{Eric Poisson} 
\affiliation{Department of Physics, University of Guelph, Guelph,
  Ontario N1G 2W1, Canada} 
\author{Roland Haas} 
\affiliation{TAPIR, Mailcode 350-17, California Institute of
  Technology, Pasadena, CA 91125, USA} 
\date{January 11, 2013} 
\begin{abstract} 
We examine Hubeny's scenario according to which a near-extremal
Reissner-Nordstr\"om black hole can absorb a charged particle and be
driven toward an over-extremal state in which the charge exceeds the
mass, signaling the destruction of the black hole. Our analysis
incorporates the particle's electromagnetic self-force and the energy
radiated to infinity in the form of electromagnetic waves. With these
essential ingredients, our sampling of the parameter space reveals no
instances of an overcharged final state, and we conjecture that the
self-force acts as a cosmic censor, preventing the destruction of a
near-extremal black hole by the absorption of a charged particle. We
argue, on the basis of the third law of black-hole mechanics, that
this conclusion is robust and should apply to attempts to overspin a
Kerr black hole.     
\end{abstract} 
\maketitle

\section{Introduction} 

The expectation that curvature singularities arising from the  
gravitational collapse of matter in general relativity should be
shielded from view by event horizons was codified in a cosmic
censorship conjecture, first formulated by Penrose in 1969
\cite{penrose:69}. While a proof is still lacking, the conjecture
is supported by numerous examples \cite{wald:97}, but it is also 
challenged by a number of potential counter-examples. Some of these
involve matter models that would be dismissed as insufficiently
physical (such as a pressureless fluid), but some are sufficiently
serious to warrant a close examination. Among these is the critical
collapse of fundamental matter fields, as investigated by Choptuik and
his collaborators \cite{choptuik:93, gundlach-martingarcia:07}; in
such cases the critical solution describes a naked singularity, but
its realization requires an initial configuration that is infinitely
finely tuned. Another is the endpoint of the Gregory-Laflamme
instability of a five-dimensional black string 
\cite{gregory-laflamme:93, lehner-pretorius:10}, which features a
horizon breaking up into ever-smaller beads joined by ever-thinner
filaments, leading to the formation of a naked singularity; but there
are no four-dimensional analogues to this instability.     

Another avenue for violating cosmic censorship was identified by
Hubeny \cite{hubeny:99}, who noticed that a near-extremal
Reissner-Nordstr\"om black hole, possessing a charge $Q$ that is
almost as large as its mass $M$ (in geometrized units in which 
$G = c = 1$), can absorb a test particle of such charge $q$, mass $m$,
and energy $E_0$ that the final configuration cannot be a black hole:
the final charge $Q + q$ exceeds the final mass $M + E_0$, signalling
the presumed destruction of the event horizon. As we shall review in
Sec.~III, Hubeny identified an open region of parameter space that
corresponds to such overcharging scenarios, revealing that they
constitute a plausible threat to cosmic censorship.  

This theme was further explored by Hod \cite{hod:02} and Jacobson and
Sotiriou \cite{jacobson-sotiriou:09}, who found that a near-extremal
Kerr black hole can absorb a test particle and be driven toward a
final state with too much angular momentum to be a black hole. In
these scenarios it is important that the initial black hole be in a
near-extremal state; an already extremal black hole would necessarily
repel the particle and prevent it from crossing the event horizon
\cite{wald:74a}. And as we shall explain in Sec.~V, it is also
important that the particle be a point particle with a vanishing
physical size \cite{hod-piran:00}: the process must be
discontinuous. We therefore exclude from our considerations attempts
to overcharge or overspin a black hole by continuous processes, for
example, by using waves instead of particles 
\cite{hod:08, kommemi:11}; such attempts will necessarily fail.  

Another important aspect of the overcharging and overspinning
scenarios is that they were analyzed on the basis of an approximate
description of the relevant physics. First, the absorbed particle was
modeled as a test particle, and all self-force, self-energy, and
radiative effects were ignored. Second, the gravitational influence of
the particle on the black-hole spacetime was not taken into account;
while one could show that the final configuration had too much charge
or angular momentum to be a black hole, the analysis could not
describe how the black hole gets dynamically destroyed. In this paper
we address the first limitation; the second limitation will not be
addressed, but our analysis indicates that the black hole will not be
destroyed by the absorption of a particle. 

Other researchers have attempted to incorporate the important 
influence of the particle's self-force, and of the radiation emitted 
during the absorption, on the overcharging and overspinning
scenarios. Hubeny, in her original work on the subject
\cite{hubeny:99}, recognized the limitations of the test-particle
analysis, and attempted to incorporate self-force effects through a
local approximation to be described in Sec.~III below. Isoyama, Sago,
and Tanaka investigated self-energy effects 
\cite{isoyama-sago-tanaka:11}, suggesting that the existence of a
turning point in the particle's motion is compatible with a final
state that is not overcharged. Barausse, Cardoso, and Khanna
incorporated the gravitational radiation emitted by a particle
attempting to overspin a Kerr black hole
\cite{barausse-cardoso-khanna:10, barausse-cardoso-khanna:11}. These
attempts were incomplete: Hubeny's self-force was approximate, Isoyama
{\it et al.}\ could not state that the self-force necessarily produces
a turning point when the black hole is about to become overcharged,
and Barausse {\it et al.}\ could not account for conservative
self-force effects.  

Our purpose in this paper is to provide a fuller analysis of the
overcharging scenarios. While our attempt is still partial (as we
shall explain in Sec.~II), it is much more complete than the ones
reviewed previously. And while these attempts could not rule out all
cases of overcharged final states, we provide evidence that when both 
conservative and dissipative aspects of the self-force are taken into
account, there are no overcharged final states. We therefore
present a case that {\it the electromagnetic self-force acts as a
  cosmic censor}, preventing the destruction of a near-extremal black
hole by the absorption of a charged particle. 

Our analysis benefits from the technical developments of the last
decade, reviewed in Ref.~\cite{poisson-pound-vega:11}, which permit
the routine computation of self-forces in curved spacetime. These
computations are relatively straightforward when the background
spacetime is spherically symmetric, and this motivates us to limit the
scope of our work to the overcharging scenarios. An analysis of the 
overspinning scenarios would require the computation of the
gravitational self-force on a particle plunging with high angular 
momentum toward a rapidly rotating Kerr black hole, and this is beyond
the current state of the art. There are also no techniques currently
available to calculate the gravitational self-force on a spinning
particle.   

\section{Electromagnetic self-force and radiated energy} 

To calculate the self-force acting on a charged particle falling 
toward a charged black hole is a formidable undertaking that is made
extremely difficult by the coupling of electromagnetic and
gravitational perturbations in the background Reissner-Nordstr\"om
spacetime. The metric is a solution to the Einstein field equations
with an energy-momentum tensor $T$ that is quadratic in the
electromagnetic field tensor $F$. The introduction of a charged
particle creates a perturbation $\delta F_1$ to the field tensor, and
a perturbation $\delta T \sim F \delta F_1$ to the energy-momentum
tensor; this produces a metric perturbation $\delta g_2$ that must be
added to the direct perturbation $\delta g_1$ created by the
particle's mass. Furthermore, the metric perturbation creates a
disturbance $\delta F_2$ in the background field tensor, which must be
added to $\delta F_1$. The perturbative problem is inherently coupled,
and techniques to calculate self-forces in such circumstances are 
not yet available. Our treatment will therefore be incomplete, in
that {\it we shall eliminate the gravitational sector} from the
perturbative problem; our electromagnetic perturbation lives in a
fixed background spacetime, and no attempt will be made to couple
it to gravity. A partial justification can be given: When the
particle's charge-to-mass ratio is very large, the gravitational
self-force associated with $\delta g_1$ can be neglected in front of
the electromagnetic self-force associated with $\delta F_1$; but the
neglect of $\delta g_2$ and $\delta F_2$ cannot be so easily
justified. We shall proceed nevertheless, and argue in Sec.~V that our 
treatment captures the essential aspects of the self-force.         

We consider a point particle falling radially toward a
Reissner-Nordstr\"om black hole. The spacetime metric is 
\begin{equation} 
ds^2 = -f\, dt^2 + f^{-1}\, dr^2 + r^2\, d\Omega^2,
\end{equation}  
in which $f = 1 - 2M/r + Q^2/r^2$ and $d\Omega^2 = d\theta^2 + 
\sin^2\theta\, d\phi^2$. The background electromagnetic field tensor 
has $F_{rt} = Q/r^2$ as its only nonvanishing component, and the
particle creates a perturbation $\delta F_{rt}$ that we decompose as 
\begin{equation} 
\delta F_{rt} = \frac{1}{r^2} \sum_{\ell m} \Phi_{\ell m}(t,r) 
Y_{\ell m}(\theta,\phi)
\end{equation} 
in terms of spherical harmonics; the other components of the
perturbation play no role in our analysis. Each mode $\Phi_{\ell m}$
of the perturbation satisfies the partial differential equation 
\begin{equation} 
-\partial_{tt} \Phi + f \partial_r (f \partial_r \Phi) 
- \frac{\ell(\ell+1)}{r^2} f \Phi = S, 
\label{eq:Maxwell} 
\end{equation} 
in which 
\begin{equation} 
S = 4\pi r^2 f ( \partial_t j_r - \partial_r j_t - 2 j_t/r) 
\end{equation} 
is a source term constructed from  
\begin{subequations} 
\begin{align} 
j_t &= -q \sqrt{\frac{2\ell+1}{4\pi}} \frac{F}{R^2} \delta(r-R) \\ 
j_r &= q \sqrt{\frac{2\ell+1}{4\pi}} \frac{\dot{R}}{F R^{2}}
 \delta(r-R), 
\end{align}
\end{subequations} 
the modes of the current density $j_\alpha$, with 
$r=R(t)$ describing the world line, $\dot{R} = dR/dt$, and 
$F = f(r=R)$. The world line is obtained by integrating the
differential equation  
\begin{equation} 
\frac{dR}{dt} = -\frac{F}{E_0 - qQ/R}\, 
\sqrt{ (E_0-qQ/R)^2 - m^2 F }.  
\label{eq:motion} 
\end{equation} 
Equation (\ref{eq:Maxwell}) is integrated numerically, making use of a
finite-difference method devised by Lousto and Price
\cite{lousto-price:97}; the method is designed to provide an exact
treatment of the delta functions on the right-hand side of
Eq.~(\ref{eq:Maxwell}).  

The integration of Eq.~(\ref{eq:Maxwell}) requires the specification
of $\Phi(t,r)$ and $\partial_t \Phi(t,r)$ at $t=0$. These initial
conditions are unknown, and in most of the self-force literature
\cite{haas:07, barack-sago:10} it has proved adequate to adopt the
trivial configuration $\Phi(0,r) = \partial_t \Phi(0,r) = 0$, in spite
of the obvious violation of the constraint equations at $t=0$. In the
usual context in which the particle moves slowly and can be followed
for a very long time, the unphysical burst of radiation that
accompanies the sudden creation of a particle at $t=0$ is of no
concern; the radiation travels away from the particle at the speed of
light, and leaves the numerical domain before the self-force is
evaluated. The present context is less forgiving. As we shall see, our
particles move extremely fast at $t=0$, and this gives little
opportunity for the radiation to peel away from the particle; and
since the particle takes little time to reach the black hole, the
numerical noise is still present when we evaluate the self-force. This
unfavorable circumstance represents a significant obstacle against
high-precision computations of the self-force. In practice we were
able to mitigate this difficulty by starting the integration when the
particle is extremely far away from the black hole, and restarting it
after some time on a smaller numerical grid, using the previously
generated results as initial data for the new run. But while this
technique does a good job at eliminating most of the noise, some
remains, and it continues to limit the accuracy of our computations
for high-speed particles.        

The electromagnetic self-force acting on the charged particle is
computed according to 
\begin{equation} 
f^r = \frac{q}{m} \biggl( E_0 - \frac{qQ}{R} \biggr) 
\delta F^{\sf R}_{rt}, 
\end{equation} 
in which $f^r$ is the radial component of the force and 
$\delta F^{\sf R}_{rt}$ is the {\it regularized} electromagnetic field
produced by the particle; this differs from the retarded solution
$\delta F_{rt}$ to Maxwell's equations by the Detweiler-Whiting
singular field \cite{detweiler-whiting:03}, which diverges at the
particle's position but is known not to contribute to the
self-force. In practice the regularized field is computed by
implementing a mode-sum regularization \cite{barack-ori:02} according  
to which 
\begin{equation} 
\delta F^{\sf R}_{rt} = \sum_{\ell} \Bigl[ (\delta F_{rt})_\ell 
- (\ell + {\textstyle \frac{1}{2}}) A - B + \cdots \Bigr], 
\end{equation} 
where $(\delta F_{rt})_\ell = r^{-2} \sum_m \Phi_{\ell m} Y_{\ell m}$ 
evaluated at the particle's position, and 
\begin{equation} 
A = \frac{q}{R^2}, \qquad  
B = \frac{qE_0}{2m R^2} - \frac{q^2Q}{mR^3} 
\end{equation} 
are regularization parameters calculated by adapting the recipe
described in Sec.~X of Ref.~\cite{casals-poisson-vega:12} to the
radial trajectories considered here. The remaining terms in the mode 
sum are given by such expressions as 
$[(\ell-\frac{1}{2})(\ell+\frac{3}{2})]^{-1}$
multiplied by additional regularization parameters; these sum to zero,
and keeping these terms accelerates the convergence of the mode sum
when it is necessarily truncated to a maximum value 
$\ell_{\rm max}$. The regularized mode sum provides a 
powerful diagnostic of numerical accuracy; the computations are 
deemed to be unreliable when $(\delta F^{\sf R}_{rt})_\ell$ fails to
fall off as $\ell^{-2}$ for large $\ell$ after subtraction of the $A$
and $B$  terms.     

The influence of the self-force on the particle's motion can be
incorporated by promoting $E_0$ to a dynamical variable $E(R)$ in
Eq.~(\ref{eq:motion}), which is related to the self-force by 
$dE/dR = \delta F_{rt}^{\sf R}$. A central question is whether the 
self-force succeeds in repelling the particle before it reaches the
event horizon; this will be the case when $E$ decreases to the extent 
that it becomes equal to $qQ/R + m\sqrt{F}$, which signals the
presence of a turning point. This occurs when 
$\delta F_{rt}^{\sf R} > 0$ and the self-force is repulsive.  

The energy radiated in the form of electromagnetic waves by the
infalling particle can also be obtained on the basis of the mode
functions $\Phi_{\ell m}$ evaluated in the limit $r \to \infty$. The
rate at which energy is radiated to infinity is given by 
\begin{equation} 
\frac{dE_\infty}{dt} = \sum_{\ell m} 
\frac{|\partial_t \Phi_{\ell m}|^2}{\ell(\ell+1)}. 
\end{equation} 
This can be integrated with respect to $t$ to obtain $E_\infty$, the
total energy radiated.    

\section{Monte Carlo sampling of overcharging trajectories} 

The self-force and the radiated energy can be computed once a choice
of trajectory $R(t)$ is made. To guide this choice we recall Hubeny's
test-particle analysis, which involves a black hole in a near-extremal
state with $Q/M = 1 - 2\epsilon^2$, where $\epsilon$ is small and
positive. The particle's charge, energy, and mass are parametrized as
\begin{equation} 
q/M = a \epsilon, \qquad 
E_0/M = a\epsilon - 2b \epsilon^2, \qquad  
m/M = c\epsilon, 
\end{equation} 
in which $(a, b, c)$ are dimensionless and of order 
unity. Hubeny showed that particles with $a > 1$, $1 < b < a$, and 
$c < \sqrt{a^2-b^2}$ produce a final configuration with 
$Q + q > M + E_0$, which is overcharged relative to a black-hole
state. This analysis ignores self-force effects, and it ignores the
energy radiated by the infalling particle. It can be shown that
incorporation of these effects does not affect the bound on $a$ nor
the upper bound on $b$. The lower bound on $b$, however, is affected
because the overcharging condition becomes  
\begin{equation} 
Q + q > M + E_0 - E_\infty
\label{eq:overcharge} 
\end{equation} 
to account for the energy radiated to infinity; since 
$q = O(\epsilon)$, $E_\infty/M$ can be parametrized as 
$\alpha \epsilon^2$ for some dimensionless quantity $\alpha$, and
$b$'s lower bound becomes $b > 1 - \frac{1}{2}\alpha$. While radiative 
effects alter the overcharging condition, the self-force determines
whether a turning point will be encountered before the particle
reaches the event horizon. To ensure that the particle's
charge-to-mass ratio is large (so that we can ignore the gravitational
self-force), we require that $c \ll a$.   

It would be hopeless to perform self-force computations for the
infinite number of trajectories that can potentially lead to
overcharging scenarios; indeed, each self-force computation requires
several hours of CPU time on a standard workstation, and each
computation must be carefully examined to ensure that it is not
vulnerable to initial-data noise. To work around these prohibitive
costs, we first performed a Monte Carlo search of the parameter space
by implementing crude approximations for the self-force and the energy
radiated. For the purposes of this search we approximated the full
self-force by a local approximation \cite{dewitt-brehme:60, hobbs:68} 
given by  
\begin{equation} 
f^\alpha_{\rm local} 
= \frac{1}{3} q^2 \bigl( g^{\alpha\beta} + u^\alpha u^\beta \bigr)
\biggl (2 \frac{Da_\beta}{d\tau} + R_{\beta\gamma} u^\gamma \biggr),
\end{equation} 
  in which $Da_\beta/d\tau$ is the particle's acceleration covariantly
differentiated with respect to proper time, and $R_{\beta\gamma}$ is
the spacetime's Ricci tensor. The flux of radiated energy is
approximated by a relativistic version of Larmor's formula, 
$dE_\infty/dt \simeq \frac{2}{3} q^2 a^\alpha a_\alpha$,
which is integrated to yield $E_\infty$. With these crude ingredients,
our Monte Carlo search involved 10,000 trajectories sampled uniformly
in the $(a, b, c)$ parametrization; the samples were taken within the
intervals $1 < a < 100$, $-200 < b < a$, $0 < c < a$, and for all
samples we set $\epsilon = 0.01$. Of these, only 27 trajectories
fulfilled the requirements for an overcharging scenario: they managed
to penetrate the black hole in spite of the repulsive action of the
self-force, and they satisfied the condition of
Eq.~(\ref{eq:overcharge}). The remaining cases were dismissed either
because the local self-force produces a turning point before the
particle reaches the event horizon, or because the Larmor formula
indicates that the overcharging condition is not satisfied.   

\section{Accurate computations} 

The Monte Carlo search was followed up with accurate computations of 
the self-force and radiated energy for a much smaller sample of
trajectories. In general we found that close to the black hole, the
actual self-force is well approximated by the local expression;
the level of discrepancy never exceeds 10\%. We also find that the
radiated energy is rather crudely approximated by the Larmor formula,
at a typical level of 40\% accuracy. 

Choosing among the cases that do not produce an overcharged final
state because of the existence of a turning point, we find that the
actual self-force tends to be larger than the local approximation,
confirming the failure of the particle to cross the event
horizon. Choosing among the cases for which the particle crosses the
horizon but Eq.~(\ref{eq:overcharge}) is not satisfied, we find that
the actual self-force also fails to produce a turning point, but the
accurate computation of $E_\infty$ confirms that the final state is
not overcharged. 

Finally, choosing among the cases that did produce an
overcharged final state (see Table~I) reveals that the actual
self-force is smaller than the local approximation (see Fig.~1), so
that it cannot succeed in producing a turning point. For these cases,
however, we find that the Larmor formula overestimates the radiated
energy, producing a final mass $M_{\rm final} = M + E_0 - E_\infty$
that is smaller than the actual value; so while $M_{\rm final}$ was
declared to be smaller than $Q_{\rm final} = Q + q$ under the Larmor 
approximation, we actually have $M_{\rm final} > Q_{\rm final}$ and a 
final state that is not overcharged. 

\begin{table}
\caption{Three sampled trajectories that were declared to produce an 
  overcharged final state in the Monte Carlo search. The table lists
  the values of $(a, b, c)$ that parametrize the choice of
  trajectory. It also specifies the initial state of the particle at 
  $t=0$; to obtain reliable results for the self-force and radiated
  energy we must begin the integrations at a very large radius, and
  with initial speeds that approach the speed of light. The table
  indicates whether the actual self-force allows the particle to cross
  the event horizon; in all cases the answer is positive. And finally,
  the table indicates whether the final state satisfies the
  overcharging condition of Eq.~(\ref{eq:overcharge}); while all
  answers would have been positive under the Larmor approximation,
  they are actually negative when the radiated energy is computed
  accurately.}
\begin{ruledtabular} 
\begin{tabular}{ccccccc} 
$a$ & $b$ & $c$ & $R/M$ & $\dot{R}$ & crossing? & overcharging? \\ 
\hline
3.728 & -46.161 & 0.7535 & 12,000 & 0.987 & yes & no \\ 
3.825 & -125.73 & 0.7829 & 20,000 & 0.992 & yes & no \\ 
3.910 & -146.10 & 0.7120 & 30,000 & 0.994 & yes & no 
\end{tabular}
\end{ruledtabular} 
\end{table}

\begin{figure} 
\includegraphics[width=3in]{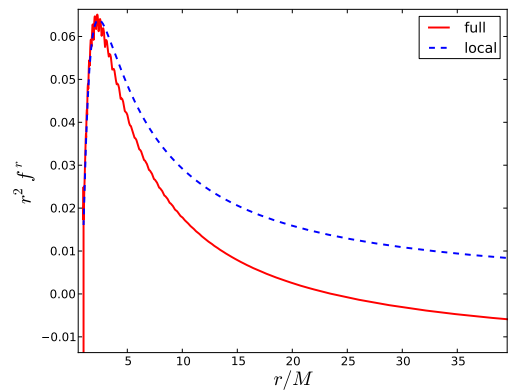}
\caption{Plot of $r^2 f^r$ as a function of $r/M$ for the first case
  listed in Table~I. The dashed line (blue online) is the local
  approximation. The solid line (red online) is the result of our
  computation. The local approximation overestimates the
  self-force except when the particle is very close to the event
  horizon, where its action is most important. The oscillations seen
  below $r/M = 5$ are a manifestation of numerical noise associated
  with an unphysical burst of radiation taking place at $t=0$.}   
\end{figure}

\section{Third law of black-hole mechanics} 

Our limited sample of the parameter space has revealed no instances of
an overcharged final state. The electromagnetic self-force seems to
act as a cosmic censor, preventing the destruction of a near-extremal
black hole by the absorption of a charged particle. To reinforce this
conclusion we elaborate an argument that suggests that it is
robust. In particular, we believe that the conclusion is not limited
by our incomplete sampling of the parameter space and our neglect of
the gravitational coupling. We believe that it would hold just as well
in attempts to overspin a Kerr black hole. The argument is based on
the third law of black-hole mechanics.      

As formulated and proved by Israel \cite{israel:86}, the third law
states that ``a nonextreme black hole cannot become extremal (i.e.\
lose its trapped surfaces) at a finite advanced time in any continuous
process in which the stress-energy tensor of accreted matter stays
bounded and satisfies the weak energy condition in a neighborhood of
the outer apparent horizon.'' (The mathematics behind the theorem 
were further developed by Andersson and his collaborators
\cite{andersson-etal:09}.)  An important aspect of Israel's theorem is 
that it is formulated in terms of the black hole's apparent horizon
and does not rely on the global existence of an event horizon (which
would be akin to assuming the validity of cosmic censorship). Another
important aspect is that the third law requires a continuous process
involving a bounded stress-energy tensor, and therefore it does not
apply to the point-particle scenarios considered here. The third law,
however, implies that any attempt to overcharge or overspin a black
hole based on continuous processes will necessarily fail.  

Our interest here is in a discontinuous process associated with a point
particle. Let us, however, consider a small but extended body that is
threatening to overcharge or overspin a nearly extremal black
hole. This body satisfies the conditions of the theorem, and it cannot
succeed in destroying the black hole; it cannot even succeed in
bringing the black hole to extremality. What is the mechanism behind
this failure?  

For the region of parameter space examined here, there would be no 
obstacle against overcharging or overspinning the black hole if the 
body were modeled as a test body in the black hole's background
spacetime. The mechanism must therefore be provided by
backreaction effects. For these overcharging or overspinning cases,
therefore, backreaction effects must force the body to turn around
before the event horizon is reached. In other words, 
{\it the net self-force acting on the extended body must provide the
  required mechanism that prevents the body from violating the third
  law.} The point particle evades the third law, but it is clear on
physical grounds that in a regime in which the extended body is
sufficiently small, {\it the self-force acting on a point particle
  will be indistinguishable from the self-force acting on the extended
  body.} And if the self-force manages to prevent the extended body
from destroying the black hole, it must do the same for the point
particle. Our conclusion, therefore, is that the self-force acts as a 
cosmic censor under all such circumstances.        

While this argument appears to us to be most plausible, we acknowledge
that it does not amount to a proof that a point particle can never be
exploited to overcharge or overspin a black hole. For example, it is
conceivable that an extended body threatening to overcharge or
overspin would break apart before reaching the horizon, with its
largest fraction turning around and a suitably small fraction being
absorbed by the black hole, keeping it in a nonextremal state. If such
a circumstance were to arise, it would be difficult to argue that the
self-force on the point particle would be indistinguishable from the
self-force on the extended body. We would, however, dismiss this 
scenario as extremely unlikely when the body is sufficiently  
small. Indeed, the destruction of the body would require strong tidal
forces (produced either by the black hole or the body's self-force),
and these will necessarily scale linearly with the body's size; while
a large body might indeed be broken up by tidal forces, a small body
will not.   

\section{Conclusion} 

We have presented two lines of argument against the destruction of a
near-extremal black hole by the absorption of a charged particle. The 
first relies on a calculation of the electromagnetic self-force and
energy radiated that neglects the gravitational sector of the
perturbation. The second relies on the third law of black-hole   
mechanics and the expectation that the motion of a point particle
cannot be distinguished from the motion of a suitably small body. Each
line of argument is incomplete. But we believe that taken together, 
they amount to a solid case in favor of the conjecture that the
self-force can act as a cosmic censor.   

\begin{acknowledgments} 
We acknowledge numerous discussions with colleagues, including Leor
Barack, Luis Lehner, Adam Pound, and Bob Wald. This work was supported
by the Natural Sciences and Engineering Research Council of Canada.  
\end{acknowledgments}    

\bibliography{../bib/master} 
\end{document}